\documentclass{iopart}
\def\d{\,{\rm d}}
\def\e{{\rm e}}
\def\sh{\mathop{\rm sh}\nolimits}

\def\Lu{\mathop\pounds_u}
\catcode`\@=11
\let\ffnsymbol\@fnsymbol
\catcode`\@=12
\eqnobysec

\begin{document}

\paper[Cylindrically symmetric perfect-fluid universes]{Cylindrically symmetric perfect-fluid universes}

\author{ P Klep\'{a}\v{c} and J Horsk\'{y}}

\address{Department of Theoretical Physics and Astrophysics, Faculty of Science,
 Masaryk University, Kotl\'{a}\v{r}sk\'{a} 2, 611 37  Brno, Czech Republic}

\begin{abstract}
The aim of this paper is to examine some obtained
exact solutions of the Einstein-Maxwell equations, especially their
properties from a chronological point of view. Each our spacetime is
stationary cylindrically symmetric and it is filled up with an
perfect fluid that is electrically charged.
There are two classes of solutions  and examples of each of them are investigated.
We give examples of the first class both for the vanishing as well as for the
non-vanishing Lorentz force. \end{abstract}

\section{Introduction}
Thanks to its remarkable properties the G\"{o}del solution (\cite{Godel},
\cite{Stephani}, \cite{Hawking}) became
subject to several interesting generalizations. Banerji and Banerjee in
\cite{Banerji} have found solutions of the Einstein-Maxwell equations for
a charged perfect fluid.  Also generalization to the case with or without the
electromagnetic field and when non-constant component ~$g_{zz}$ or 
$g_{tt}$ of the metric tensor is admited is possible (\cite{Harvey} and
references citated therein). In this paper we give a number of solutions of
the Einstein-Maxwell equations with the vanishing cosmological constant
which are generalizations of the metrics discussed in~\cite{Banerji} and
which involve electromagnetic field as well as the non-constancy of
$g_{zz}$ or $g_{tt}$. These solutions result from two classes of
exact solutions of the Einstein-Maxwell equations for a spacetime filled
up with the charged perfect fluid.
The subject of this work directly
generalizes the paper given by Mitskievi\v{c} and Tsalakou~\cite{Mit}, who
used the H-M conjecture (\cite{Horsky 89}), and therefore
their designations are partly preserved.  

The plan of the paper is as follows. In section~\ref{sec1} we obtain the
Einstein-Maxwell equations for the stationary and cylindrically
symmetric spacetime with vector
potential (\ref{rce2}). Through the whole paper we systematically
use an orthonormal basis in which the stress-energy tensor of the perfect
fluid is diagonal. In section~\ref{sec2} we get a general
solution of the first class that corresponds to the non-constant
component~$g_{zz}$ of the metric tensor. In the case of vanishing Lorentz
force we write down the inequalities implied by the energy conditions as well
as the values of the Riemann tensor for this solution. Section~\ref{sec4}
derives a second class corresponding to the
non-constancy of $g_{tt}$ component of the metric tensor.

\newpage

\section{The Einstein-Maxwell equations\label{sec1}}

In this paper, we consider a spacetime filled up with the charged perfect
fluid of the pressure $p$ and the energy density $\mu$. Let us choose
local coordinate system $x^\mu =
(t,\varphi,z,r)$ \footnote[3]{The coordinates will be numbered
$(t,\varphi,z,r)=(0,1,2,3)$.} of the comoving coordinates, where $\varphi$
is angular, $r$ radial and $z$~ordinary Cartesian coordinates.
We search for the metric which is stationary and
cylindrically symmetric and
depends on the only coordinate $r$, so we will restrict our
attention to the case when the spacetime admits Killing vector fields in
remaining directions, i.e. in $t,\ \varphi$ and $z$~directions. Since we
demand the metric to be invariant over simultaneous time reversion
$t\rightarrow -t$ and reflection $\varphi\rightarrow -\varphi$,
and over the inversion $z\rightarrow -z$, the
spacetime metric tensor has the form \footnote[4]{We use the  units with
the speed of light $c$ and the Newtonian gravity constant $G$ equaled to one.}
\begin{equation}
\d s^2=\e^{2\alpha}\left(\d
t+f\d\varphi\right)^2-l^2\d\varphi^2-\e^{2\gamma}\d
z^2-\e^{2\delta}\d r^2\label{rce1}\ ,
\end{equation}
where all metric functions as well as the pressure and the charge
density depend only on the coordinate~$r$. The function $\delta $ can
achieve any value. The velocity vector is given by
$u = \e^{-\alpha}\partial_t$ and the acceleration of the particles of
the fluid is
\begin{equation}
\dot{u}=\nabla_{\!\!u} u=\alpha^\prime e^{-2\delta}\partial_3\ ,\label{zrychleni}
\end{equation}
where comma means derivative with respect to~$r$. The expansion tensor as well
the shear tensor of the spacetime with the metric (\ref{rce1}) is vanishing
\[ \Theta_{\mu\nu}=\frac12\Lu h_{\mu\nu}=\frac12\Lu
g_{\mu\nu}-\dot{u}_{(\mu}u_{\nu)}=0\ .\]
Finally, let the one-form of the vector
potential be expressed like (see \cite{Bonnor})
\begin{equation}
{\rm A}=m(r)\d\varphi+n(r)\d t \ .
\label{rce2}
\end{equation}
Let us introduce an orthonormal basis with the basic one-forms as follows
\begin{equation}
\eqalign{\Theta^{\hat{0}} = \e^\alpha(\d t+f\d\varphi)\ ,  \qquad
& \Theta^{\hat{1}} = l\d\varphi\ , \\
\Theta^{\hat{2}} = \e^\gamma\d z\ ,&  \Theta^{\hat{3}} = \e^\delta\d r\ .}
\label{rce5}
\end{equation}

Using~(\ref{rce2}) for mixed components of the two-form of the 
electromagnetic field one gets
\begin{equation}\label{Lorentz}
\fl F_\mu^{\ \nu}=l^{-2}\delta^3_\mu\left\{\left[n^\prime
l^2e^{-2\alpha}+(m^\prime-fn^\prime)f\right]\delta_0^\nu
-(m^\prime-fn^\prime)\delta_1^\nu\right\}+\e^{-2\delta}\delta^\nu_3
(n^\prime\delta^0_\mu+m^\prime\delta^1_\mu).
\end{equation}
The Maxwell equations with sources $\delta F =4\pi j $ give
\begin{equation}
\fl l^{-1}\e^{-\alpha-\gamma-\delta}\{\e^{\alpha+\gamma-\delta}l^{-1}
[[n^\prime l^2\e^{-2\alpha}+(m^\prime-fn^\prime)f]\delta^\mu_0
-(m^\prime-fn^\prime)\delta^\mu_1]\}_{,3}=-4\pi j^\mu,\label{rce3}
\end{equation}
$j(r)$ being the current density, $\delta j=0$. We postulate that the
fluid particles carry the charge, $j=\rho u$, $\rho$ is the charge
density. It will be accomplished if
\begin{equation}
(m^\prime - fn^\prime)\e^{\alpha+\gamma-\delta}=Bl\ ,
\label{rce4}
\end{equation}
$B$ is constant. The equations (\ref{rce3}) and (\ref{rce4}) result in the
following expression for the charge density
\begin{equation}\label{charge}
4\pi \rho =-\frac{B^2}{m^\prime-fn^\prime}\frac {\d}{\d r}\left (\frac{l^2n^\prime \e^{-2\alpha }}{m^\prime -fn^\prime }
+ f\right )\e^{-\alpha-2\gamma }.
\end{equation}
\noindent The choice (\ref{rce2}) corresponds to the formulae for the
electric and the magnetic fields~(\cite{Mit})  
\begin{equation}
{\rm E}_{\hat{r}}=n'\e^{-\alpha -\delta }\ ,\ {\rm B}_{\hat{z}}=B\e^{-\alpha-\gamma}.
\label{intenz}
\end{equation}
With the help of the constraint (\ref{rce4}) the vorticity one-form is given by
\[\omega=\frac12\ast(u\wedge\d u)=\frac{Bf^\prime}{2\left(m^\prime-fn^\prime \right)}\d z\ .\]

The total stress-energy tensor of the spacetime $T_{total} $ arises as a sum 
of the stress-energy tensor of the electromagnetic field
\begin{eqnarray} 
\fl T_{elmag}  & =(8\pi)^{-1}l^{-2}\e^{-2\delta}\{\ 2n^\prime (m^\prime -fn^\prime )l\e^{-\alpha}(\Theta^{\hat{0}}\otimes\Theta^{\hat{1}}
+\Theta^{\hat{1}}\otimes\Theta^{\hat{0}}) \\ \fl  & +\left[n^{\prime 2}l^2\e^{-2\alpha}+(m^\prime-fn^\prime)^2\right]\Theta^{\hat{0}}\otimes\Theta^{\hat{0}}  +\left[n^{\prime 2}l^2\e^{-2\alpha}+(m'-fn')^2\right]\Theta^{\hat{1}}\otimes\Theta^{\hat{1}} \nonumber \\
\fl & +\left[n^{\prime 2}l^2\e^{-2\alpha}-(m^\prime-fn^\prime)^2\right]\Theta^{\hat{2}}\otimes\Theta^{\hat{2}} 
-\left[n^{\prime 2}l^2\e^{-2\alpha}-(m^\prime-fn^\prime)^2\right]\Theta^{\hat{3}}\otimes\Theta^{\hat{3}} \}  \nonumber \label{elmag}
\end{eqnarray}
and the stress-energy tensor of the perfect fluid
\begin{equation} \label{fluid}
T_{fluid}=\mu\Theta^{\hat{0}}\otimes\Theta^{\hat{0}}
+p\left(\Theta^{\hat{1}}\otimes\Theta^{\hat{1}} +
\Theta^{\hat{2}}\otimes\Theta^{\hat{2}}+
\Theta^{\hat{3}}\otimes\Theta^{\hat{3}}\right)\ .
\end{equation}
The Bianchi identities reduce to the single equation of the motion of the
fluid

\begin{equation}\label{Bianchi}
p^\prime +\alpha^\prime\left(p+\mu\right)+\rho n^\prime =0\ .
\end{equation}

The non-zero components of the Einstein tensor in the orthonormal basis
(\ref{rce5}) can be written in the form
\begin{eqnarray}
\fl G_{\hat{t}\hat{\varphi}}  =R_{\hat{t}\hat{\varphi}}=\frac12\ \e^{-2\alpha-\gamma-\delta}\frac{\d}{\d
r}\left(l^{-1}f^\prime
\e^{2\alpha}\e^{\alpha+\gamma-\delta}\right)\ ,\ \nonumber \\
\fl G_{\hat{\varphi}\hat{\varphi}}-G_{\hat{r}\hat{r}} =l\e^{-\alpha-\gamma-\delta}\frac{\d}{\d r}
\left(l^{-1}(\alpha+\gamma)^\prime
\e^{\alpha+\gamma-\delta}\right)-2\alpha^\prime\gamma^\prime
\e^{-2\delta}\ ,\ \nonumber \\ \fl G_{\hat{r}\hat{r}}+G_{\hat{z}\hat{z}} =\frac{1}{2}\ l^{-1}\e^{-\alpha-\gamma-\delta}
\frac{\d}{\d r}\left[l^{-1}\e^{-\alpha+\gamma-\delta}\frac{\d}{\d r}\left(l^2\e^{2\alpha}\right)\right] \ ,\ \nonumber \\
\fl G_{\hat{t}\hat{t}}-G_{\hat{r}\hat{r}} =-\frac{1}{2}\ l^{-1}\e^{-\alpha-\gamma-\delta}\frac{\d}{\d
 r}\left[l^{-1}\e^{\alpha-\gamma-\delta}\frac{\d}{\d r}\left(l^2\e^{2\gamma}\right)\right]+\frac12l^{-2}f^{\prime
2}\e^{2(\alpha-\delta)}\ ,\ \nonumber \\ \fl G_{\hat{r}\hat{r}} =\left(\alpha^\prime
l^\prime+l^\prime\gamma^\prime+l\gamma^\prime\alpha^\prime\right)l^{-1}\e^{-2\delta}
+\frac14l^{-2}f^{\prime 2}\e^{2(\alpha-\delta)}\ ,\nonumber \end{eqnarray}
and the Einstein-Maxwell equations $G_{\hat{\mu}\hat{\nu}}= 8\pi T_{\hat{\mu}\hat{\nu}}$ are
\numparts\begin{eqnarray}
\fl \frac{\d}{\d r}\left[\frac{\e^{2\alpha}f^\prime}{m^\prime-fn^\prime}\right]=4n^\prime\ ,\ \label{Einstein}\\
\fl \frac{\d}{\d r}\left[\frac{(\alpha+\gamma)^\prime}{m^\prime-fn^\prime}\right]-
\frac{2\alpha^\prime\gamma^\prime}{m^\prime-fn^\prime}=
\frac{2n^{\prime 2}\e^{-2\alpha}}{m^\prime-fn^\prime}\ ,\ \\
\fl \frac{1}{2}\frac{\d}{\d r}\left[\frac{\e^{-2\alpha}}{m^\prime-fn^\prime}\frac{\d}{\d r}\left(l^2\e^{2\alpha}\right)\right]
=\frac{16\pi  l^2p \, \e^{2\delta}}{m^\prime-fn^\prime}\ ,\ \label{tlak} \\
\fl \frac{1}{2}\frac{\d}{\d  r}\left[\frac{\e^{-2\gamma}}{m^\prime-fn^\prime}
\frac{\d}{\d r}\left(l^2\e^{2\gamma}\right)\right]=\frac{8\pi(p-\mu)}{m^\prime-fn^\prime}l^2\e^{2\delta}
+\frac12\frac{f^{\prime2}\e^{2\alpha}}{m^\prime-fn^\prime}-
\frac{2n^{\prime 2}l^2 \e^{-2\alpha}}{m^\prime-fn^\prime}\ ,\ \label{Bhust} \\
\fl (\alpha^\prime+\gamma^\prime)\frac{\d \, l^2}{\d r}+2\alpha^\prime\gamma^\prime l^2
=16\pi l^2p \, \e^{2\delta}-\frac12 f^{\prime 2}\e^{2\alpha}-
2n^{\prime 2}l^2\e^{-2\alpha}+2(m^\prime-fn^\prime)^2 \ .\label{Einstein2}
\end{eqnarray}           \endnumparts

In summary, we obtained the system of six equations
\eref{Einstein}-\eref{Einstein2} and \eref{rce3} minus
\eref{Einstein2}, which is an integral of the \eref{Bianchi}. After
eliminating of $p$ from \eref{Einstein}-\eref{Bhust} with the help of
the \eref{Einstein2} one gets the system of six equations for totally
nine unknown functions $\alpha, \gamma,
\delta, f, l, \mu, \rho, m$ and $n.$ It means that three out of nine
functions can be chosen arbitrarily. These three degrees of freedom
are equivalent to introducing $\delta,\ \rho$ and $\mu(p).$
In the next we will consider the $m$
and $n$ fixed, and moreover we impose, as in \cite{Mit}, aditional
condition: either $\alpha$ or $\gamma$ are constant, which is in accordance
with two classes of solutions.

\section{First class -- $\alpha$ is constant\label{sec2}}

When $\alpha ={\rm const}$ the Einstein-Maxwell equations
(\ref{Einstein})-\eref{Bhust}, \eref{rce3} can be
easily integrated in terms of the electric and the magnetic potential
$m$ and $n$. \Eref{Bianchi} now says that the Lorentz force is
balanced by the pressure gradient, ${\rm grad} p+\rho {\rm
E}=0$, resulting in the geodesic motion of
the fluid particles.
We can, without loss of generality, suppose that $\alpha$
is zero. The metric components are
\numparts       \label{pok}
\begin{eqnarray}f\e^{2n^2}=4\ \int m^\prime n\e^{2n^2}\d r+F \ ,\label{rce6} \\
\gamma =\int\left[\int\frac{2n^{\prime 2}}{m^\prime-fn^\prime}\d r+C\right]
(m^\prime-fn^\prime)\d r \ ,\   \\
l^2\e^{-2\gamma}=E
-4\ \int{\left(m-fn+k\right)(m^\prime-fn^\prime)\e ^{-2\gamma}\d r}.
\label{urc}
\end{eqnarray} \endnumparts
In (\ref{rce6}) $\ C,\ E,\ F,\ k$ are constants of integration.
Inserting \eref{rce6} into \eref{Einstein2} and \eref{Bhust} yields for the
pressure and the energy density
\begin{equation}\label{Bpres}
\eqalign{
 8\pi p=B^2\left[4n^2+\frac{\gamma^{\prime
  2}+n^{\prime 2}}{\left(m^\prime-fn^\prime\right)^2}l^2
  -2\gamma^\prime\frac{m-fn+k}{m^\prime-fn^\prime}-1\right]\e^{-2\gamma},\\
 8\pi\mu=B^2\left[4n^2-\frac{3\gamma^{\prime
  2}+5n^{\prime 2}}{\left(m^\prime-fn^\prime\right)^2}l^2
 +6\gamma^\prime\frac{m-fn+k}{m^\prime-fn^\prime}+1\right]\e^{-2\gamma}.}
\end{equation}

\subsection{Example of the first class with non-vanishing Lorentz force\label{subssec3.1}}

We give here explicitly the solution when $ m=\tau n+\frac23\beta n^3$, $\tau$
and $\beta$ being constants, and $k,\ F$ and $C$ in
\eref{rce6}-\eref{urc} are zero.
In this case (according to the (\ref{Lorentz}) the Lorentz force does not vanish)
the metric can be written like this
\begin{eqnarray}             \label{noncon}
\fl \d s^2&=\left[\d t+\left(2\beta n^2+\tau-\beta\right)\d \varphi\right]^2
-\frac{\beta^2}{3}\left(3E\beta^{-2}\e^{2n^2}-4n^2+1\right)\d
\varphi^2\nonumber\\ \fl &-\e^{2n^2}\d z^2-\frac{1}{B^2}\frac{3n^{\prime
2}\e^{2n^2}\d r^2}{3E\beta^{-2}\e^{2n^2}-4n^2+1}\ ,\
\end{eqnarray}
and for the charge density (\ref{charge}), for the the pressure and the
energy density \eref{Bpres} one has
\begin{equation} \label{hustnon}
\eqalign{\fl 8\pi p=B^2\left [\frac{E}{\beta^2}\left (4n^2+1\right )
-\frac{2}{3}\e^{-2n^2}\right ] \ ,\
\pi \rho = -B^2n\left(\frac{E}{\beta^2}+\frac{1}{3}\e^{-2n^2}\right)\ , \\
\fl 8\pi \mu =B^2\left\{\left[\frac{56}{3}n^2-\frac{2}{3} \right]\e^{-2n^2}
-\frac{E}{\beta ^2}\left(12n^2+5 \right) \right\}.}
\end{equation}   

\subsection{Solution with the vanishing Lorentz force\label{subssec3.2}}

\noindent Metric given by (\ref{rce6})-\eref{urc} is not too transparent.
In the rest of this section we restrict ourselves to purely magnetic field
when the electric potential $n$ is constant, which will be marked as $\frac14b$,
i.e. when according to (\ref{Lorentz}) or \eref{rce4} the Lorenz
force $F^\mu_{\ \nu}u^\nu$ vanishes.
The result written in terms of the magnetic potential $m$
is \footnote[5]{In fact, there is another way of
deriving localy the same metric as (\ref{rce7})
that introduces new variable by rescalling the coordinate~$r$ by
the definition $m(r)=m$. Then for the two form of the electromagnetic
field $F=\d m\wedge\d\varphi$. In effect, this substitution
results in inserting unity except~$m^\prime$.}($F$~was~gauged~away)
\begin{eqnarray} 
\d s^2 & =\left(\d t+bm\d\varphi\right)^2-(E\e^{2Cm}+\lambda
m+\nu)\d\varphi^2-\e^{2Cm}\d z^2 \nonumber \\
& -\frac{1}{B^2}\frac{m^{\prime 2}\e^{2Cm}}{E\e^{2Cm}+\lambda m+\nu}\d r^2\ , \label{rce7}\\
& \lambda =\frac{4-b^2}{2C}\ ,\ \nu =\frac{\lambda+4k}{2C}\ . \nonumber
\end{eqnarray}
\label {lambda}The charge density, pressure and the energy density are
\begin{equation}\label{rce8}
\eqalign{4\pi\rho &= -B^2b\e^{-2Cm}\ ,\ 8\pi p=B^2C^2E\ ,\\ 8\pi\mu &=
B^2\left[(b^2-2)\e^{-2Cm}-3C^2E\right]\ .}
\end{equation}

Our fluid moves geodesically, without shear and expansion \label{lambda}
\footnote[6]{Thanks to the homogeneity of the pressure (\ref{rce8}),the
solution (\ref{rce7}) can be reinterpreted also as a solution describing
charged dust in the spacetime with the cosmological constant
$\Lambda =-B^2C^2E$ (the Einstein equations in our convention are
$ G-\Lambda g=8\pi T$).\label {foot} }.
For $m$ being increasing or decreasing function defined on
$\left(0,\infty\right)$ and covering the whole range $\left(0,\infty\right)$,
it can be treated as a new radial variable and one obtains the first family of the solutions in \cite{Mit}, 
if we interpret the coordinate $\varphi$ as an ordinary Cartesian coordinate (i.e. one abandons the periodicity of the $\varphi$ coordinate). 
The same will be true in the section \ref{sec4}. The solution without the electromagnetic field ($B=0$) was recovered by Wright (\cite{Wright}).

\subsubsection { Energy conditions.\label{subsubsec3.2.1}}
\numparts
Dominant energy condition implies inequalities
\begin{eqnarray}
(b^2-2)\e^{-2Cm}-2C^2E\geq 0\ ,\label{rce10}\\
(b^2-2)\e^{-2Cm}-4C^2E\geq 0\ .\label{rce11} 
\end{eqnarray}\endnumparts
In our case the strong energy condition is satisfied if the dominant energy condition is. 
Altogether the energy conditions will be fulfiled if (\ref{rce10}) or (\ref{rce11}) is satisfied depending 
on whether $E$ is negative or positive respectively, or $b^2 \geq 2$, if $E$ is zero.

\subsubsection{Curvature.\label{subsubsec3.2.3}}

\noindent For the future convenience we mention the concrete values of
the Riemann curvature tensor 
for the metric~(\ref{rce7}). All non-vanishing components of the
Riemann tensor are given by
\begin{equation}\label{sing}
\eqalign{R_{\hat{r}\hat{z}\hat{r}\hat{z}}
=R_{\hat{\varphi}\hat{z}\hat{\varphi}\hat{z}}=
\frac12 B^2\e^{-2\alpha}\left(\lambda
C\e^{-2Cm}+2C^2E\right)\ ,\  \\
R_{\hat{t}\hat{r}\hat{t}\hat{r}}
=R_{\hat{t}\hat{\varphi}\hat{t}\hat{\varphi}}=
-\frac14 B^2b^2\e^{-4\alpha}\e^{-2Cm} \ ,\ \\
R_{\hat{t}\hat{r}\hat{\varphi}\hat{r}}
=-R_{\hat{t}\hat{z}\hat{\varphi}\hat{z}}=
\frac12 B^2bC\e^{-3\alpha}(E\e^{2Cm}+\lambda
m+\nu)^{\frac12}\e^{-2Cm} \ ,\ \\
R_{\hat{\varphi}\hat{r}\hat{\varphi}\hat{r}}
=-\frac14B^2\e^{-2\alpha}(3b^2\e^{-2\alpha}+2\lambda
C)\e^{-2Cm}+B^2C^2E\e^{-2\alpha}\ . }
\end{equation}

\subsubsection{Closed timelike curves.\label{subsubsec3.2.2}}

\noindent For the analysis of potential chronology violation we use
the Carter theorem~(\cite{Carter}, \cite{Tipler}). In our case (\ref{rce7})
the isometry group, generated by the Killing vector fields
$\partial_t,\ \partial_\varphi,\ \partial_z$, is Abelian and
timewise orthogonally transitive, since the hypersurfaces of
transitivity given by constant~$r$ are everywhere timelike,
$g^{rr}<0$. The only exception could be the case $C=0$, but then the
chronological structure follows from the continuity and we will meet
such an example in subsubsection \ref{sec3}.
Chronology will be preserved when
one is able to find such constants $p, q, s$ that the
differential form $\psi = p\d t + q\d\varphi + s\d z$
is everywhere timelike or null. Since the contribution of $s$ is always negative,
it can be treated as zero.

\subsubsection{Example of the first class with vanishing Lorentz force.\label{sec3}}

Let the vector potential be ${\rm A}=2a^2B\sh^2(r/2a)\d \varphi$. Here $a$ is a length characteristic\footnote[7]
{With respect to the previous footnote, $a^2=-\left(2\Lambda \right)^{-1}$ holds 
in the alternative reinterpretation.}, $B$ is according to (\ref{intenz})
 a value of the magnetic field on the
rotation axis. The choice of the constant of integration
$E=-\nu=(2a^2B^2C^2)^{-1}\ ,\ b^2=4-4CB^{-1}+2\left(aB\right)^{-2}$
yields the metric
\begin{eqnarray}
\fl \d s^2 &=\left[\d t+2a\left(4a^2B^2+2-2C\right)^{\frac12}m\d \varphi\right]^2-
\frac{2a^2}{C^2}\left[\e^{2Cm}-2C(1-C)m-1\right]\d \varphi^2\nonumber \\
\fl &-\e^{2Cm}\d z^2-\frac{2a^2C^2m^{\prime 2}\e^{2Cm}\d r^2}{\e^{2Cm}-2C(1-C)m-1}\ ,\
\label{rce17}
\end{eqnarray}
where, for convenience, we have denoted  
$2a^2BC\rightarrow C$ and $m=\sh^2(r/2a)$. For the physical 
quantities $\mu$, $p$ and $\rho$ one obtains
\begin{equation}\label{rce18}
\eqalign{8\pi\mu=2\left(B^2+\frac{1-C}{a^2}\right)\e^{-2Cm}-\frac{3}{2a^2}\ ,\\
2\pi\rho=-B\left(B^2+\frac{1-C}{2a^2}\right)^{\frac12}\e^{-2Cm}\
\ ,\ 8\pi p=\frac{1}{2 a^2}\ .}
\end{equation}

Signature in~(\ref{rce17}) is always correct and the energy
condition~(\ref{rce11}) requires $C\leq 0$.
From equations (\ref{sing}) we can see that for negative~$C$ the
physical singularity occurs when $r\rightarrow\infty$. From (\ref{rce17})
one can also see that no event horizon is present for any finite value of $r$.
In the limit~$C=0$ we get Banerji-Banerjee solution (formulae (11) and (12) in
the alternative interpretation) in \cite{Banerji} which is spacetime homogeneous
and singularity free. So we can interpret~$C$ as an indicator of the difference
of the spacetime from the spacetime homogeneity.

In fact, the choice $m=\sh^2(r/2a)$ represents the family of models in which $m$
is increasing or decreasing function with domain $\left(0,\infty\right)$ and
image $\left(0,\infty\right)$. Each such choice after the introducting of the
new variable $m=m\left(r\right)$ leads to the \eref{rce17}.
From this point of view it is obvious that for example the choice
$m\propto r^{-1}$ has the same physical content as
$m=\sh^2(r/2a)$. The same remark will be true in subsection \ref{sec5}.

The condition $\psi_\alpha\psi^\alpha \geq 0$ of the subsubsection
\ref{subsubsec3.2.2} reads
\begin{equation}\label{rce19}
\fl \frac{1}{2C^2}[e^{2Cm}-2C(1-C)m-1]-\left[\left(4a^2B^2+2-2C\right)^{\frac12}m
-\frac{q}{2ap}\right]^2\geq 0\ .
\end{equation}
Because $q$ must be zero in order not to predominate in
(\ref{rce19}) for small $r$, the condition for the non-existence of the
closed timelike curves (hereafter~CTC)
has the familiar form $g_{\varphi\varphi}\leq 0$ (or $g_{\varphi\varphi} <
0$ for the non-existence of the closed causal curves), and we
see that CTC will apper always for sufficiently large $r$.

\begin{table}
\caption{Numerical dates that relate
the quantity of the magnetic field~$B$ on the rotation axis to the radius $R$ of
the first null circle $(t,\ z,\ r={\rm const} )$, and to the cosmological
constant~$\Lambda$ in the alternative interpretation.
For simplicity, dates are given only for $C=0$. 
The matter density used is~$10^{-26}\mbox{kg}\cdot\mbox{m}^{-3}$, 
which corresponds to the period of the rotation $7\cdot 10^{10}\, \mbox{years}$.}
\begin{indented}
\item\begin{tabular}[t]{@{}lr@{}lr@{}lr@{}lr@{}lr@{}lr@{}l}
\br
$B\ [10^{3}\mbox{Gauss}]$ & 0&,0 & 3&,7 & 4&,8 & 6&,7 & 8&,2 & 10&,5 \\
\mr
$\Lambda\ [10^{-53}\mbox{m}^{-2}]$ & -9&,31 & -8&,20 & -7&,45 & -5&,59
& -3&,73 & -0&,28 \\
$R\ [10^8\mbox{ly}]$ & 1&37 & 1&32 & 1&29 & 1&23 & 1&18 & 1&10\\
\br
\end{tabular}
\end{indented}
\label{tab1} \end{table}

Thanks to absence of any event horizons, for any two spacetime points $p$ and $q$
, one has $p\gg q$ and simultanously $q\gg p$ (\cite{Hawking}).
Especially $I^+(p)=I^-(p)=J^+(p)=J^-(p)~=~M$. Although
through every point passes some CTC, each such curve must
inevitable cross the region
$r>R$, where $R$ is determined as a root of~(\ref{rce19}).
The CTC we are interested in are nontrivial (\cite{Carter}).
Moreover, because the Carter theorem can be applied independently on
whether or not $\varphi$ is periodic, one can see that CTC occur even if
we interpret $\varphi$ as the ordinary Cartesian coordinate.
(The chronology could be preserved in the case that we took only part of the
(\ref{rce17}) confined to the region $r<R$ and tried to match it to some other
chronologically well behaved solution.)

The magnetic field has always positive influence on the
chronology violation, precisely the larger value of the magnetic field
on the rotation axis the closer to it will be the "first" null circle followed
by CTC (see table~\tref{tab1}).

Similarly, another Banerjee-Banerji solution (expression (15) in
\cite{Banerji}) or the Som and Raychaudhuri solution \cite{Bonnor} can
be got by suitable choice of the constant of integration.

\section{Second class of the solutions -- $\gamma$ is constant}\label{sec4}

Solution with constant~$\gamma$ will be refered as second class and here we
restricted ourselves only to the case when the Lorentz force vanishes,
with the electric potential $n=b/4$. \Eref{urc} remains valid in this case
too. $\gamma$ can be put
to zero and the solution of the Einstein-Maxwell equations
(\ref{Einstein})-\eref{Bhust} and \eref{rce3} can be written in the form
\begin{equation}\label{rce24}
\eqalign{f=-\frac{b}{2C}\e^{-2Cm}+H\ ,\e^{2\alpha}=\e^{2Cm} \ , \\
l^2=\frac{b^2}{4C^2}\e^{-2Cm}-2m^2+Dm+E\ ,}
\end{equation}
with constants of integration $\ C,\ H$ and $D=Hb-4k$. For
the energy density, the pressure and the charge density one has
\begin{equation}\label{rce25}
\eqalign{16\pi\mu=B^2\e^{-2Cm}[CD+2(1-2Cm)]\ , \\
16\pi p=B^2\e^{-2Cm}[CD-2(1+2Cm)]\ ,\
4\pi\rho=-B^2b\e^{-3Cm}.}
\end{equation}
Thanks to inhomogeneity of the pressure (\ref{rce25}), the fluid already
does not move along geodesic lines. Its acceleration (\ref{zrychleni}) is
\[\nabla_{\!\!u} u=C m^\prime e^{-2\delta}\partial_3\ .\]
 
In contrary with the corresponding solution of the first class (\ref{rce17}), 
because of the inhomogeneity of the pressure, the metrics of the second class 
cannot be reinterpreted as describing dust with the non-zero cosmological
constant.

\subsection{Energy conditions\label{subssec5.1}}

In the case of the second class (\ref{rce24}) and (\ref{rce25}) the energy
conditions altogether will be fulfilled if following inequality will hold
\begin{equation}\label{rce26}
CD\geq 4Cm\ .
\end{equation}

\subsection{Curvature\label{subssec5.2}}

The non-zero components of the Riemann tensor are given by
\begin{equation}\label{rce27}
\eqalign{R_{\hat{t}\hat{r}\hat{t}\hat{r}}
=R_{\hat{t}\hat{\varphi}\hat{t}\hat{\varphi}}=
B^2C\e^{-2\alpha}\left(2m-\frac{D}{2}\right)\e^{-2Cm}\ ,\ \\
R_{\hat{r}\hat{\varphi}\hat{r}\hat{\varphi}}
=B^2\e^{-2\alpha}\left[C\left(2m-\frac{D}{2}\right)-2\right]\e^{-2Cm}\ . }
\end{equation}

\subsection{Example of the second class of the solutions\label{sec5}}

This special solution corresponds to that  discussed
in~\ref{sec3} in the sense that when $C$ goes to zero, we
obtain the same solution as in~\ref{sec3} when $C$ goes to
zero, i.e. (11) and (12) in \cite{Banerji}. For let us choose
${\rm A}=2a^2B\sh^2(r/2a)\d \varphi$ and constants
of integration $b^2=4+2\left(aB\right)^{-2},D=2B^{-1}+2C^{-1}
+\left(a^2B^2C\right)^{-1},E=-b^2\left(2C\right)^{-2},H=b\left(2C\right)^{-1}.$
Components of the metric tensor read (denoting $2a^2BC\rightarrow C$ and
$m=\sh^2(r/2a)$)
\begin{equation}\label{rce28}
\eqalign{\e^{2\alpha}=\e^{2Cm}\ ,\
f=a\frac{\left(4a^2B^2+2\right)^{\frac12}}{C}\left(1-\e^{-2Cm}\right)\ , \\
\fl l^2=4a^2\left[\frac{2a^2B^2+1}{2C^2}\e^{-2Cm}-2a^2B^2m^2+\frac{(2a^2B^2+1+C)m}{C}
-\frac{2a^2B^2+1}{2C^2}\right]\ .}
\end{equation}
The energy density, the pressure and the charge density are given respectively
by
\begin{eqnarray}\label{hust6.1}
8\pi\mu=\left(2B^2+\frac{1+C}{2a^2}-2B^2Cm\right)\e^{-2Cm}\ , \nonumber \\
8\pi p=\left(\frac{1+C}{2a^2}-2B^2Cm\right)\e^{-2Cm}\ ,\\
2\pi\rho=-B\left(B^2+\frac{1}{2a^2}\right)^{\frac{1}{2}}\e^{-3Cm}\ .\nonumber
\end{eqnarray}

It can be seen from the form of metrics (\ref{rce28}) that the signature is
correct for $C\leq 0$ and, in case when~$B$ is zero, then for every~$C$.
The absence of  the event horizons for any finite value of $r$ is apparent
from the form of the metric (\ref{rce28}).
From analysis of the energy conditions~(\ref{rce26}) it follows that
these will be fulfilled for intervals $-(2a^2 B^2+1)\leq C\leq 0$, 
and if the electromagnetic field vanishes for $C\geq -1$. 
One can convince oneself that CTC will occur
if $-(2a^2B^2+1)\leq C\leq 0$ when $B\neq 0$, and $C > -1$ when $B=0$. Every two
points of the spacetime can be connected each other by both future
and past directed timelike curve which is non-trivial, so that the spacetime is
totally vicious (\cite{Carter}). Formulae (\ref{rce27})
show that the physical singularity occurs for $C<0$ when $r$ tends to the
infinity. The only exception is the case with the vanishing 
electromagnetic field ($B=0$) and $C=-1$, which is after transformation
$ \e ^{-2m}=1-\omega^2u^2$, with $ \omega^{-2}=2a^2$, the Minkowski spacetime
in the rotating cylindrical coordinates $\varphi\rightarrow \varphi - \omega t$.

\section{Conclusion} 

From the requirement of the stationarity and cylindrical
symmetry and from the assumption that the spacetime is filled up
with an perfect fluid that is charged, we have obtained the system
of ordinary differential equations (\ref{Einstein})-\eref{Bhust} plus
\eref{rce3}, that was solved in
terms of the functions $m$ and $n$
in two cases: the first class~(\ref{rce6})-\eref{urc}, \eref{Bpres}
with constant~$\alpha$ and
the second class~(\ref{rce24}), (\ref{rce25}) with constant~$\gamma$
(but we gave the solution of the second class only for the vanishing
Lorentz force). These two general classes are both shear-free.
Fluid of the first class moves geodesically and its velocity is the
Killing vector field. It is not true in the second class thanks to the
inhomogeneity of the pressure (\ref{rce25}). 
We shown explicit special solution of the first class for the non-vanishing
Lorentz force (formulae (\ref{noncon}) and (\ref{hustnon})), but without
detailed analysis, and in some details we discussed the example of the
first class (subsubsection \ref{sec3}) and the corresponding example of the
second class (subsection \ref{sec5}) with the vanishing Lorentz force, that
are generalizations to the solution given in \cite{Banerji}. These
metrics with three parameters (the length characteristic $a$,
the value of the magnetic field on the rotation axis $B$, and~$C$
determining the difference of the solutions from the homogeneity)
contain CTC at least for some interval of the
values of~$C$. It turns out that the magnetic field has positive
influence on the appearance of the non-trivial chronology violation (in fact by
sufficiently large magnetic field we can always ensure the
chronology violating). 

\ack {The authors thank to Professor M. J. Reboucas and to referees for their advices and stimulating
comments.}

\end{document}